\documentclass{ws-procs975x65}
\newcommand{\ba}{\begin{eqnarray}}
\newcommand{\ea}{\end{eqnarray}}

\begin{document}

\title{Critical Points in Nuclei and\\
   Interacting Boson Model Intrinsic States}

\author{Joseph N. Ginocchio}

\address{Theoretical Division, Los Alamos National Laboratory,
Los Alamos, NM, 87545, U.S.A.\\E-mail: gino@lanl.gov}

\author{ A. Leviatan}

\address{Racah Institute of Physics, The Hebrew University,
Jerusalem 91904, Israel\\
E-mail: ami@vms.huji.ac.il}

\maketitle

\abstracts{
We consider properties of critical points in the interacting
boson model, corresponding to flat-bottomed potentials
as encountered in a second-order phase transition between spherical
and deformed $\gamma$-unstable nuclei. We show that intrinsic states
with an effective $\beta$-deformation reproduce the dynamics of the
underlying non-rigid shapes. The effective deformation can be determined
from the the global minimum of the energy surface after projection onto
the appropriate symmetry. States of fixed $N$ and good $O(5)$ symmetry
projected from these intrinsic states provide good analytic estimates
to the exact eigenstates, energies and quadrupole transition rates at
the critical point.}
      ``{\it During these moments of abstraction he seemed more intimately
absolved, in the sense of  being linked anew with the universe. }

Giuseppe de Lampedusa, ``The Leopard"

\section{Introduction}

In the days that one of us (JNG) was a graduate student, group theory
was considered almost a
dirty word in the nuclear physics community even though Wigner had
introduced spin-isospin
symmetry (SU(4)), Elliott had exploited the
symmetry of the
harmonic oscillator to link
collective motion and the
shell model (SU(3)) and the symmetry of the
hadrons had been
discovered
(SU(3) again). Francesco Iachello changed that attitude
and brought group
theory front and center in nuclear physics and in
other fields of physics.

Franco Iachello is a descendant of a noble
Sicilian family similar to
that portrayed in Giuseppe de Lampedusa's
classic Italian novel,
``The Leopard".
Set in Sicily in 1860 at the
time of the campaign for the unification of
Italy, the hero, the
Prince, struggles with how to keep the old while
embracing the new.
The Prince often took solace from the turmoil of the
real world by
studying astronomy and mathematics much the same as Franco
has by his
significant contributions to physics and group
theory.

\section{Critical Points in the Geometric Collective
Model}

Recently Franco has been studying two critical points
associated with
shape phase transitions in nuclei within the
geometric
framework of the collective model for infinite square well
potentials
\cite{iac00,iac01} . This model involves a Bohr
Hamiltonian
which describes the dynamics of a macroscopic quadrupole
shape
via a differential equation in the intrinsic quadrupole shape
variables
beta and gamma. In the current contribution we shall
discuss the
critical point (CP) for a second order shape phase
transition
between spherical and deformed $\gamma$-unstable
nuclei,
which Franco called E(5) \cite{iac00}.
An empirical example
of such a critical point
has been found in
$^{134}$Ba~\cite{cas00,arias01} and possibly
in
$^{104}$Ru~\cite{frank02}, $^{102}$Pd~\cite{zamfir02}
and
$^{108}$Pd~\cite{zhang02}.

\begin{table}[t]
\caption{Excitation
energies (normalized to the energy of the
first excited state) and
B(E2) values (in units of
$B(E2;\, 2^{+}_{1,1}\rightarrow
0^{+}_{1,0})=1$)
for the E(5) critical point
{\protect\cite{iac00}},
for several N=5 calculations and for the
experimental data
of $^{134}$Ba {\protect\cite{ba134}}.
The finite-N
calculations involve
the exact diagonalization of the critical
IBM
Hamiltonian ($H_{cri}$) [Eq.~(\ref{hcri})], $\tau$-projected
states
for $H_{cri}$ [Eqs.~(\ref{enexit}),(\ref{be2xit}) with
$y=0.314$],
the $U(5)$ limit [$\epsilon\,n_d$] and the $O(6)$
limit
[$(A/4)(N-\sigma)(N+\sigma+4)+B\tau(\tau+3)$].}
\begin{center}
\footnotesize
\begin{tabular}{lcccccc}
\hline
&
E(5) & exact   &
$\tau$-projection & U(5) & O(6) & $^{134}$Ba \\
&  &
N=5 & N=5   & N=5  & N=5  & exp  \\
\hline
$E(0^{+}_{1,0})$ & 0    &
0     & 0      & 0  & 0 & 0       \\
$E(2^{+}_{1,1})$ & 1    & 1
& 1      & 1  & 1 & 1       \\
$E(L^{+}_{1,2})$ & 2.20 & 2.195 & 2.19
& 2  & 2.5 & 2.32  \\
$E(L^{+}_{1,3})$ & 3.59 & 3.55  & 3.535  & 3  &
4.5 & 3.66  \\
$E(0^{+}_{2,0})$ & 3.03 & 3.68  & 3.71   & 2
&
$1.5{A\over B}$ & 3.57  \\
\hline
$B(E2;\, 4^{+}_{1,2}\rightarrow
2^{+}_{1,1})$ &
1.68 & 1.38 & 1.35 & 1.6 & 1.27 & 1.56(18)
\\
$B(E2;\, 6^{+}_{1,3}\rightarrow 4^{+}_{1,2})$ &
2.21 & 1.40 & 1.38
& 1.8 & 1.22 &  \\
$B(E2;\, 0^{+}_{2,0}\rightarrow 2^{+}_{1,1})$
&
0.86 & 0.51 & 0.43 & 1.6 & 0 & 0.42(12)
\\
\hline
\end{tabular}
\end{center}
\end{table}

In the geometric
approach
the E(5) eigenfunctions \cite{iac00}
are proportional
to
Bessel functions of order $\tau + {3\over 2}$
and the
corresponding eigenvalues are proportional to
$(x_{\xi,\tau})^2$.
Hamiltonians that are $\gamma$- unstable have an
$O(5)$ symmetry and
$\tau$ is the $O(5)$ quantum number. Furthermore

$x_{\xi,\tau}$ is the $\xi$-th root of these Bessel functions.
A
portion of an E(5)-like spectrum is shown in Fig.~1. It consists
of
states, $L^{+}_{\xi,\tau}$, arranged in major families labeled
by
$\xi=1,2,\ldots$ and $O(5)$ $\tau$-multiplets
($\tau=0,1,\ldots$)
within each family. The angular momenta $L$ for
each $\tau$-multiplet
are obtained by the usual $O(5)\supset O(3)$
reduction \cite{ibmo6}.
The E(5) CP leads to analytic
parameter-free
predictions for energy ratios and $B(E2)$ ratios
which
persist when carried over to a finite-depth potential
\cite{caprio02}.
As seen in Table 1, the E(5) predicted values
are
in-between the values expected of a spherical vibrator [$U(5)$]
and a
deformed $\gamma$-unstable rotor
[$O(6)$].

\begin{figure}[t]
\begin{center}
\epsfxsize=25pc
\epsfbox{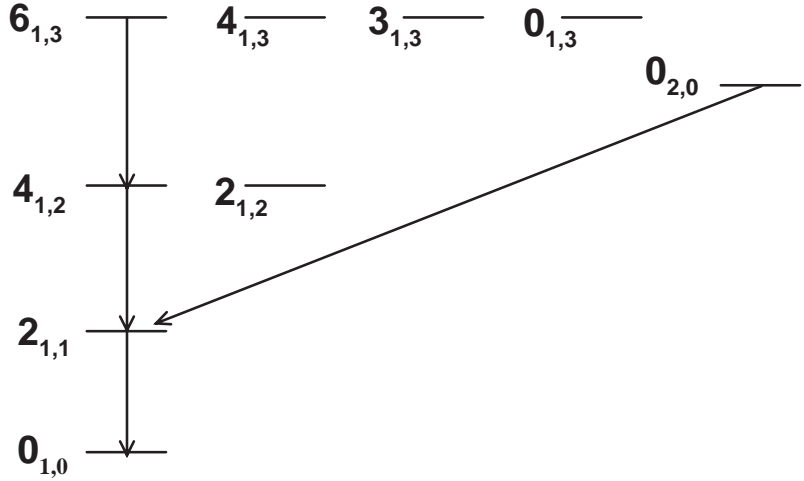}
\caption{An
E(5)-like spectrum of states labeled by $L_{\xi,\tau}$.
Shown are the
transitions whose B(E2) values are given in
Eq.~(\ref{be2xit}). The
E2 rates for other
$\Delta\tau=1$ transitions (not shown) are
governed by O(5) symmetry.
Specifically, $B(E2;\,
L^{+}_{1,2}\longrightarrow 2^{+}_{1,1})$
for $L=4,2$ are in the ratio
$1:1$ respectively,
$B(E2;\, L^{+}_{1,3}\longrightarrow 4^{+}_{1,2})$
for $L=6,4,3$
and $B(E2;\, L^{+}_{1,3}\longrightarrow 2^{+}_{1,2})$
for $L=4,3,0$
are in the ratios $1: 10/21: 2/7: 11/21: 5/7:
1$,
respectively. Taken from ref.{\protect\cite{levgin03}}.
\label{E5}}
\end{center}
\end{figure}

\section{Wave Function Ansatz
for the $\gamma$-Unstable Limit of the
Interacting Boson Model}

The
Interacting Boson Model \cite{ibm} (IBM) was one of the
great
achievements of Akito
Arima and Franco Iachello.
This model
describes low-lying quadrupole collective states in nuclei
in terms
of a system of $N$ monopole ($s$) and
quadrupole ($d$) bosons
representing valence nucleon pairs.
The IBM Hamiltonian relevant to
the critical point of the phase transition
between spherical and
$\gamma$-unstable deformed nuclei has $O(5)$ symmetry
\cite{diep80}.
Its energy surface, obtained by the method of coherent
states
\cite{diep80,gino80}, is $\gamma$-independent and exhibits
a
flat-bottomed behavior in $\beta$ which resembles the infinite
square-well
potential used to derive the E(5) CP in the geometric
approach
\cite{iac00}. Calculations with finite $N$ values ($N$=5 for
$^{134}$Ba)
have found that this critical IBM Hamiltonian
can
replicate numerically the E(5) CP and
its analytic predictions
\cite{cas00,frank02,zamfir02,zhang02}.
In the present contribution,
based on recent work \cite{levgin03},
we examine the properties and
conditions
that enable features of E(5) CP to occur in a finite
system described by
the interacting boson model. For that purpose
we
propose wave functions of a particular analytic form,
which can
simulate accurately the exact IBM eigenstates at the critical
point.
These wave functions with fixed $N$ and good $O(5)$
symmetry are used
to derive accurate estimates for energies and
quadrupole rates at the
critical point without invoking large-$N$
approximations. The
proposed wave functions can be obtained by projection
from intrinsic
states with an effective $\beta$-deformation.

In the IBM, the $\gamma$-unstable transition region is modeled by
the Hamiltonian
\ba
H &=& \epsilon\,\hat{n}_d + {1\over 4}A\,
\left[\, d^{\dagger}\cdot d^{\dagger} -  (s^{\dagger})^2\,\right ]
\left[\, H.c.\,\right]
\label{hamilt}
\ea
with $\epsilon$ and $A$ positive parameters. Here $\hat{n}_d$ is the
$d$-boson number operator, $H.c.$ stands for Hermitian conjugate and
the dot implies a scalar product.
In the $U(5)$ limit ($A=0$), the spectrum of $H$ is harmonic,
$\epsilon\, n_d$, with $n_d=0,1,2\ldots N$. The eigenstates are
classified according to the chain
$U(6)\supset U(5)\supset O(5) \supset O(3)$ with quantum numbers
$\vert\,N,n_d,\tau,L\rangle$ (for $\tau\geq 6$ an additional multiplicity
index is required for complete classification).
These states can be organized into sets characterized by $n_d=\tau
+2k$.
States in the lowest-energy set
($k=0$)
satisfy
\ba
&&P_0\,\vert\,N,n_d=\tau,\tau,L\rangle =
0
\nonumber\\
&&P_0^{\dagger}=d^{\dagger}\cdot d^{\dagger}
~.
\ea
Other sets ($k >0$) are generated by
$\vert\,N,n_d,\tau,L\rangle
\propto
(P^{\dagger}_0)^{k}\vert\,N-2k,n_d=\tau,\tau,L\rangle$.
In the $O(6)$ limit ($\epsilon=0$), the spectrum is
${1\over 4}A(N-\sigma)(N+\sigma+4)$ with
$\sigma=N,N-2,N-4,\ldots 0\; {\rm or}\; 1$.
The eigenstates are classified according to the chain
$U(6)\supset O(6)\supset O(5) \supset O(3)$ with quantum numbers
$\vert\,N,\sigma,\tau,L\rangle$. The ground band has
$\sigma = N$ and its members satisfy
\ba
&&P_1\,\vert\,N,\sigma=N,\tau,L\rangle =
0
\nonumber\\
&&P^{\dagger}_1 =
[\, d^{\dagger}\cdot d^{\dagger} -
\,
(s^{\dagger})^2\,] ~.
\ea
The remaining bands with $\sigma=N-2k$
are generated by
$\vert\,N,\sigma,\tau,L\rangle
\propto
(P^{\dagger}_1))^{k}\,\vert\,N-2k,\sigma
,\tau,L\rangle.$
These results suggest that in-between the $U(5)$ and
$O(6)$ limits,
we consider a ground band ($\xi=1$) for the
Hamiltonian
({\ref{hamilt}})
determined by
the
condition
\ba
&&P_y\,\vert\,
\xi=1;y,N,\tau,L\rangle =
0
~,
\nonumber\\
&&P^{\dagger}_y
=
\left[\,
d^{\dagger}\cdot
d^{\dagger} - y\, (s^{\dagger})^2\,\right
].
\label{p0}
\ea
In the
$U(5)$ basis these states
are
\ba
\vert\,
\xi=1;y,N,\tau, L\rangle
&=&
\sum_{n_d}{1\over
2}\left [1 +
(-1)^{n_d-\tau}\right ]\,
\xi_{n_d,\tau}
\vert\,
N,n_d,\tau, L\rangle ~,
\label{projt}
\ea
and
the $n_d$ summation covers the range
$\tau\leq n_d\leq N$.
The
coefficients $\xi_{n_d,\tau}$ have the
explicit
form
\ba
\xi_{n_d,\tau} &=& \left
[
(N-\tau)!\,(2\tau+3)!!\over
(N-n_d)!\,(n_d-\tau)!!\,(n_d+\tau+3)!!\,\right
]^{1/2}\,
y^{(n_d-\tau)/2}\,\xi_{\tau,\tau}
~,
\nonumber\\
\left
(\xi_{\tau,\tau}\right )^2 &=& {2(N-\tau+1)\over
(2\tau+3)!!}\,
y^{2\tau+3}\,\left [ G^{(\tau+1)}_{N+1-\tau}(y)\right
]^{-1} ~ ,
\nonumber\\
G^{(n)}_{\alpha}(y) &=& 2 y^{2n +
1}\sum_{p}
\left({\alpha\atop 2p + 1}\right )
\, y^{2p}\,
{(2p+1)!!\over (2p + 2n + 1)!!}
~.
\label{xindt}
\ea
$G^{(n)}_{\alpha}(y)$ is an odd function of $y$,
$G^{(n)}_{\alpha}(-y) = -G^{(n)}_{\alpha}(y)$, and satisfies
the
follwing recursion relation
\ba
G^{(n)}_{\alpha}(y)
&=&
{1\over
\alpha+2}\,G^{(n-1)}_{\alpha+2}(y)
-{1\over \alpha+1}\,G^{(n-1)}_{\alpha+1}(y) \quad\quad n\geq 1
\nonumber\\
G^{(0)}_{\alpha}(y) &=& (1+y)^{\alpha} - (1-y)^{\alpha} ~.
\ea
Furthermore, $G^{(n)}_{\alpha}(y) =
\pm {2^{\alpha +
n}\,\alpha(\alpha+n-1)!\over (\alpha + 2n)!}$
for $y = \pm 1$ and
$G^{(n)}_{\alpha}(y) \sim  {2\alpha\over (2n+1)!!}\,y^{2n+1}$
for $y \to 0$.

Members of the first excited band $(\xi=2)$ have
approximate wave functions of the form
\ba
&&\vert\, \xi=2; y,
N,\tau, L\rangle =
{\cal N}_{\beta}\,P^{\dagger}_{y}\,\vert\, \xi=1;
y,N-2,\tau,L\rangle
\nonumber\\
&&{\cal N}_{\beta} = \left [ \,
2(2N+y^2 +1) +
4(y^2-1)S^{(N-2)}_{1,\tau}\, \right ] ^{-1/2} ~,
\label{projtbeta}
\ea
where $S^{(N)}_{1,\tau}$ is defined in
Eq.~(\ref{s1k}) below.

The states of Eqs.~(\ref{projt}) and
(\ref{projtbeta}) have fixed $N$,
$L$ and good $O(5)$ symmetry
$\tau$.
Henceforth, for reasons to be explained below, they will be
referred to as $\tau$-projected states. Diagonal matrix elements of
the Hamiltonian (\ref{hamilt}) in these states, denoted by
$E_{\xi,\tau}=
\langle \xi; y, N, \tau\vert H \vert \xi; y, N,
\tau\rangle $,
can be evaluated in closed form
\ba
&&E_{\xi=1,\tau} =
\epsilon\left[\, N - S^{(N)}_{1,\tau}\,\right ]
+ {1\over 4}A\,
(1-y)^2\, S^{(N)}_{2,\tau}
\nonumber\\
&&E_{\xi=2,\tau} =
\epsilon\left\{ N - 2{\cal N}^{2}_{\beta}\,
\left [\, 2y^2 + (2N + 7y^2 -1)\,S^{(N-2)}_{1,\tau}
+ 2(y^2-1)\,S^{(N-2)}_{2,\tau}\,\right ]\,\right\}
\nonumber\\
&&\;\;\; +{1\over 4}A\,2{\cal N}^{2}_{\beta}\,
\left\{\, 2(y-1)^2\,(y^2-1)\,S^{(N-2)}_{3,\tau}
+ (y-1)^2\,(2N+y^2 -8y + 5)\,S^{(N-2)}_{2,\tau}\right.
\nonumber\\
&&\;\;\; \left.
\qquad\qquad\quad
+ 16(y-1)(N+y)\,S^{(N-2)}_{1,\tau}
+ 2\left [\,(2N+y)(2N+y+2) + 1\,\right ]\right\} ~.
\qquad
\label{enexit}
\ea
The $E_{\xi,\tau}$
are independent of $L$
since
$H$ is an $O(5)$ scalar. Their
expressions involve
the
quantities
$S^{(N)}_{k,\tau} =\langle \xi=1;
y,N,\tau\vert
(s^{\dagger})^ks^k
\vert \xi=1;y, N, \tau\rangle$ which
are given by
\begin{eqnarray}
S^{(N)}_{1,\tau} &=&
(N-\tau + 1)
{G^{(\tau+1)}_{N-\tau}(y)\over
G^{(\tau+1)}_{N-\tau+1}(y)}
\label{s1k}
\end{eqnarray}
with
$S^{(N)}_{2,\tau}
= S^{(N)}_{1,\tau}S^{(N-1)}_{1,\tau}$
and
$S^{(N)}_{3,\tau}
=
S^{(N)}_{1,\tau}S^{(N-1)}_{1,\tau}S^{(N-2)}_{1,\tau}$.
Non-diagonal
matrix
elements of the Hamiltonian $H$
between $\tau$-projected states
in
different $\xi$-bands, $H_{1,2;\tau}=
\langle \xi=2; y, N,
\tau\vert
H \vert \xi=1; y, N, \tau\rangle$,
can be evaluated as
well
\begin{eqnarray}
&&H_{1,2;\tau} =
\nonumber\\
&&\quad
2{\cal
N}_{\beta}\left[S^{(N)}_{2,\tau}\right]^{1/2}
\left\{\, \epsilon\,y
+
{1\over 4}A(y-1)\left[\, (2N+y+1) +
2(y-1)\,
S^{(N-2)}_{1,\tau}\,\right]\right\}
~.\quad
\label{nondiag}
\end{eqnarray}
By techniques
similar to that
employed in the $O(6)$ limit of the IBM
\cite{ibmo6},
explicit
expressions of quadrupole rates can be derived for
transitions
between the $\tau$-projected states. For the relevant
IBM quadrupole
operator, $T(E2) = d^{\dagger}s +
s^{\dagger}\tilde{d}$,
these
transitions are subject to the $O(5)$
selection rule
$\Delta\tau=\pm 1$,
and, as explained in the caption
of Fig.~1, it is
sufficient to focus
on the B(E2) values
\ba
&&B(E2;\, \xi=1;\,
\tau+1, L=2\tau+2
\longrightarrow
\xi=1,\,
\tau,L=2\tau)
=
\nonumber\\
&&\quad\qquad
{(\tau+1)\over
(2\tau+5)(N-\tau+1)}\,
\left
(
S^{(N)}_{1,\tau}\right
)^2\,
{G^{(\tau+1)}_{N-\tau+1}(y)\over
G^{(\tau+2)}_{N-\tau}(y)}
\left
[
\, y
+
(N-\tau)
{G^{(\tau+2)}_{N-\tau-1}(y)\over
G^{(\tau+1)}_{N-\tau}(y)}\right
]^2
~,
\nonumber\\
&&B(E2;\,
\xi=2,\,
\tau,
L=2\tau\longrightarrow
\xi=1,\,
\tau+1,L=2\tau+2)
=
\nonumber\\
&&\quad\qquad
{(\tau+1)(4\tau+5)\over
(4\tau+1)(2\tau+5)}
4\,{\cal
N}^{2}_{\beta}\,
y^2\,(y-1)^2\,
(N-\tau)\,
{G^{(\tau+1)}_{N-\tau-1}(y)\over
G^{(\tau+2)}_{N-\tau}(y)}
~.
\label{be2xit}
\ea

\section{Intrinsic
State of
the
$\gamma$-Unstable Wavefunction Ansatz}

The states
in
Eqs.
(\ref{projt}) and Eq.~(\ref{projtbeta}) can be obtained
by
$O(5)$
projection from the IBM intrinsic states for the
ground
band
\begin{eqnarray}
\vert\,c;
N
\rangle
&=&
(N!)^{-1/2}(b^{\dagger}_{c})^N\,\vert
0\,\rangle
\nonumber\\
b^{\dagger}_{c}
&=&
(1+\beta^2)^{-1/2}\left
[\, \beta\,\cos\gamma\,
d^{\dagger}_{0}
+
\beta\,\sin{\gamma}\,
{1\over\sqrt{2}}\left (
d^{\dagger}_{2}
+
d^{\dagger}_{-2}\right )
+
s^{\dagger}\,\right]
\label{cond}
\end{eqnarray}
and for the
$\beta$
band respectively
\begin{eqnarray}
\vert\,\beta; N
\rangle
&=&
{\cal
N}_{\beta}P^{\dagger}_{y}\vert\, c;
N-2\rangle
~,
\label{betaint}
\end{eqnarray}
provided $y=\beta^2$.
The
expressions in Eqs.~(5)-(12)
depend on the so far
unspecified
parameter $y$.
Normally, the equilibrium value
of
$\beta$, and hence
$y$, is chosen as
the global minimum of the
intrinsic energy surface

determined from the expectation value of $H$
in the intrinsic state
(\ref{cond}).
This is a standard procedure for a
Hamiltonian
describing nuclei with rigid shapes, for
which the global
minimum is deep and well-localized.
Such is the case for the
Hamiltonian of Eq.~(\ref{hamilt}) whose
intrinsic energy surface has
the form
\ba
E(\beta) &=& E_0
+
N(N-1)\,\beta^2(1+\beta^2)^{-2}\,
\left [ \,a + c\,\beta^2 \,
\right ]
\nonumber\\
a &=& \bar{\epsilon} - A \;\;, \;\; c =
\bar{\epsilon}
\;\; ,\;\;
\bar{\epsilon} =
\epsilon/(N-1)
\label{enegen}
\ea
with $E_0 =
\frac{1}{4}AN(N-1)$ a
constant.
The topology of the energy surface
is such
that
\ba
\begin{array}{lll}
a > 0 & & \;\;\, {\rm
global\;\,
spherical\;\, minimum\;\, at\;\,} \beta=0\\
a < 0 & &
\;\;\, {\rm
global\;\, deformed\;\, minimum\;\, at\;\,}
\beta =
a/(a-2c) ~.\\
\end{array}
\ea
When $a\neq 0$ the intrinsic energy
surface
behaves quadratically ($\sim \beta^2$)
near the single
minimum, and
the standard procedure of determining the
equilibrium
value of $\beta$ from the global minimum of
$E(\beta)$ is applicable.
However, near the critical point of the phase transition
an alternative procedure is required.

\section{$\gamma$-unstable
Wave
function Ansatz at the Critical Point}

The IBM Hamiltonian,
$H_{cri}$, at the critical point of the
$U(5)-O(6)$ phase
transition
corresponds to a special
choice of parameters in the
Hamiltonian of
Eq.~(\ref{hamilt})
\begin{equation}
H_{cri}:\quad
\epsilon = (N-1)A ~,
\label{hcri}
\end{equation}
for which $a=0$ in
Eq.~(\ref{enegen}), and the corresponding
energy surface reduces
to
\ba
E(\beta) &=& E_0 +
A\,N(N-1)\,\beta^4(1+\beta^2)^{-2}
~.
\label{enesurf}
\ea
In this case, the intrinsic energy surface
$E(\beta)$,
shown in Fig.~(2a), has a flat
behavior ($\sim \beta^4$) for small $\beta$, an inflection
point at $\beta=1$ and approaches a constant for large $\beta$.
The global minimum at $\beta=0$ is not well-localized
and $E(\beta)$ exhibits considerable instability in $\beta$,
resembling a square-well potential for $0\leq\beta\leq 1$.
Under such circumstances fluctuations in $\beta$ are large and
play a significant role in the dynamics. Some of their effect can
be taken into account by introducing into the intrinsic states
of Eqs.~(\ref{cond}) and (\ref{betaint}) an effective
$\beta$-deformation.
The effective deformation is expected to be in the range
$0 < y=\beta^2 < 1$, in-between the respective $U(5)$ and
$O(6)$ value of $\beta$. This will enable a reproduction of E(5)
characteristic signatures which are in-between these limits (see
Table I). In contrast to $E(\beta)$ of Eq.~(\ref{enesurf}),
we see from Fig.~(2b) that the
$O(5)$ projected energy surface
$E_{\xi=1,\tau=0}(y)$ of $H_{cri}$
(Eqs.~(\ref{enexit}) and
(\ref{hcri}) with $N=5$),
does have a stable minimum at a certain value of $y$,
which we interpret as an effective $\beta$-deformation.
This procedure, based on variation after
projection, is in the spirit of \cite{otsuka87} in which it is
shown that in finite boson systems, a $\gamma$-unstable $O(6)$ state
can be generated from a rigid triaxial intrinsic state with an
effective $\gamma$-deformation of 30$^{\circ}$.
In the present case
the $\gamma$-instability is treated exactly by means of $O(5)$
symmetry, while the $\beta$-instability is treated by
means of an effective deformation.
The appropriate value of $y$ can be used to
evaluate the band-mixing,
$\eta_{\tau}(y)={|H_{1,2;\tau}|\over E_{2,\tau}- E_{1,\tau}}$.
A small value of $\eta_{\tau}$ will ensure
that the $\tau$-projected
states of
Eqs.~(\ref{projt}),
(\ref{projtbeta}) form a good
representation of
the actual eigenstates of
$H_{cri}$, and turn the
expressions of
Eqs.~(\ref{enexit}), (\ref{be2xit})
into meaningful
estimates for
energies and quadrupole transition rates at
the
critical
point.

\begin{figure}[t]
\begin{center}
\epsfxsize=25pc
\epsfbox{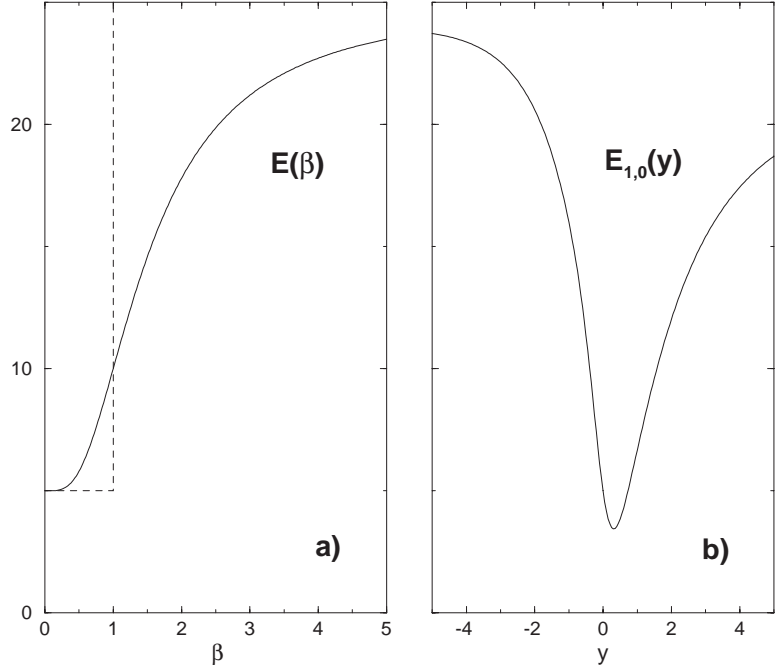}
\caption{Energy
surfaces
of
the critical IBM Hamiltonian $H_{cri}$
(\ref{hcri}) with
$N=5$ and
$A=1$.
(a)~Intrinsic energy surface
$E(\beta)$,
Eq.~(\ref{enesurf})
[solid line], and its
approximation
by a square-well potential
[dashed line].
(b)~$O(5)$ projected energy
surface
$E_{\xi=1,\tau=0}(y)$,
Eq.~(\ref{enexit}).
The global minimum
is at
$y=0.314$.
Taken
from
ref.{\protect\cite{levgin03}}.
\label
{energy}}
\end{center}
\end{figure}

\section{Comparison
with E(5) and Experiment}

To test the suggested procedure we compare
in Table 2
the $U(5)$ decomposition of exact eigenstates
obtained
from numerical diagonalization
of $H_{cri}$ for $N=5$ with
that calculated
from the $\tau$-projected states
with $y=0.314$ [the
global minimum
of $E_{1,0}(y)$].
As can be seen, the latter provide a
good
approximation to
the exact eigenstates (the
corresponding
band-mixing
is
$\eta_{\tau}=0.12,\,3.53,\,4.14,\,3.05\%$ for
$\tau=0,1,2,3$).
This
agreement in the structure of wave functions is
translated also
into
an agreement in energies and B(E2) values as
shown in Table 1.
The
results of Table 1 and 2 clearly demonstrate the ability
of the
suggested procedure to provide analytic and accurate
estimates
to the
exact finite-N calculations of the critical IBM
Hamiltonian,
which
in-turn agree with the experimental data in
$^{134}$Ba and
captures the
essential features of the E(5)
critical
point.

\begin{table}[t]
\caption[]{$U(5)$
decomposition (in \%) of
the $L^{+}_{\xi,\tau}$ states
for $N=5$.
The calculated values are
obtained from the $\tau$-projected states,
Eqs.~(\ref{projt}),
(\ref{projtbeta}) with $y=0.314$.
The exact
values are obtained from
numerical diagonalization of the
critical
IBM Hamiltonian
$H_{cri}$,
Eq.~(\ref{hcri}).}
\begin{center}
\footnotesize
\begin{tabular}{|r|cc|cc|cc|cc|cc|}
\hline
&\multicolumn{2}{|c|}{$0^{+}_{1,0}$}
&
\multicolumn{2}{c|}{$2^{+}_{1,1}$}
&
\multicolumn{2}{c|}{$L^{+}_{1,2}$}
&
\multicolumn{2}{c|}{$L^{+}_{1,3}$}
&
\multicolumn{2}{c|}{$0^{+}_{2,0}$}
\\
$n_d$
&
calc & exact & calc & exact &
calc & exact &
calc & exact & calc
&
exact \\
\hline
0 & 83.2 & 83.4
&      &      &      &      &
&
& 15.8 & 16.4 \\
1 &      &
& 92.2 & 90.2 &      &      &      &
&
&      \\
2 & 16.4 & 16.2
&      &      & 96.8 & 95.2 &      &
& 70.9
& 76.2 \\
3 &      &
& 7.8  & 9.7  &      &      & 99.1 &
98.4 &
&      \\
4 &  0.4 &
0.4 &      &      & 3.2  & 4.8  &
&      & 13.3
&  7.4 \\
5 &      &
& 0.0  & 0.1  &      &      &
0.9 &  1.6
&
&
\\
\hline
\end{tabular}
\end{center}
\end{table}

\section{Large
$N$ Limit}

In the large $N$ limit, using Stirling's Formula
in
Eq.~(\ref{xindt}),
we obtain
\ba
G^{(n)}_{\alpha}(y) \rightarrow
2
x\, \left ({x\over N}
\right)^{2n}\,\sum_{p}\
\,{
x^{2p}\over
(2p)!!(2p + 2n + 1)!!} ~,
\label{Glarge}
\ea
where $x =
Ny$.
Therefore,
\ba
S^{(N)}_{k,\tau}(y) \rightarrow {\cal
S}_{k,\tau}(x)
~,
\label{Slarge}
\ea
that is,
$S^{(N)}_{k,\tau}(y)$
becomes a function of $x$ only. This suggests
that the energy
spectrum depends only on $x$ and not
$N$ and $y$
separately as $N$ goes to infinity,
except for possibly an overall
scale. To test this in Fig.~3
we plot the energy ratio at the
critical point
\ba
R ={E_{\xi=1,\tau= 2}
-E_{\xi=1,\tau= 0}
\over
E_{\xi=1,\tau=1} - E_{\xi=1,\tau= 0}}
\label{ratio}
\ea
for several
values of $N$ as a function of $x$.
Indeed as $N$ increases, the curves asymptote
to one universal curve. This means that the quadrupole beta
deformation of the geometrical model is
proportional to $\sqrt{N}\beta$. This observation may be useful in
relating the predictions of E(5) CP to the IBM.

\begin{figure}[t]
\begin{center}
\epsfxsize=25pc
\epsfbox{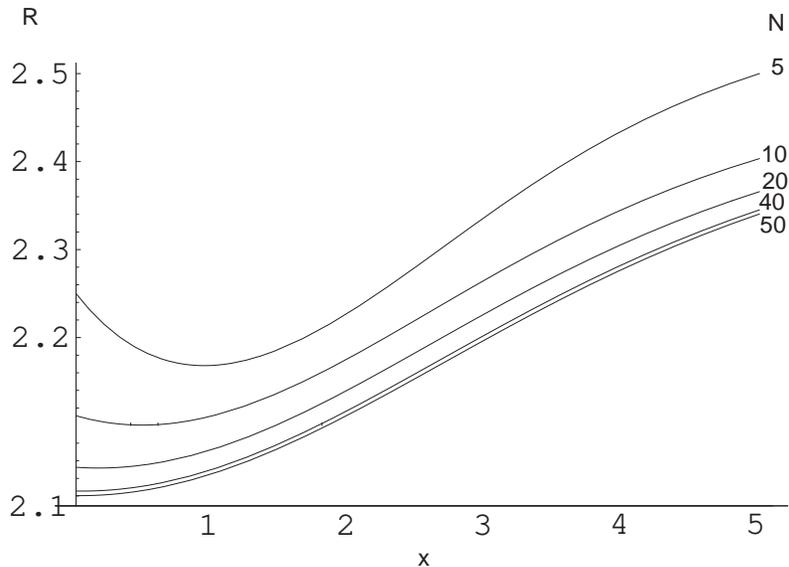}
\caption{The
energy ratio $R$ at the critical point defined in
Eq. (\ref {ratio})
as a function of $x=Ny$, for different values of N.
Note that the
vertical axis is displaced from
zero.}
\label{R}
\end{center}
\end{figure}

\section{Summary and
Future Outlook}

In this contribution we have examined properties of
a critical point in a finite system\cite{levgin03}.
We have focused on the E(5) critical point
relevant to a second-order shape phase transition between spherical
and deformed $\gamma$-unstable nuclei.
At the critical point of such a phase
transition the intrinsic energy surface is flat and
there is no stable minimum value of the deformation.
However, for a finite system, we have shown that there is an effective
deformation which can describe the dynamics at the critical point.
The effective deformation is determined by minimizing the energy surface
after projection onto the appropriate symmetries. States of finite $N$
and good $O(5)$ symmetry, projected from intrinsic states with
this effective deformation simulate accurately the exact eigenstates,
and can be used to derive analytic estimates for energies and quadrupole
transition rates at the critical point.

In the future we shall explore in depth the $N$-dependence as well
as the large-$N$ limit and its relationship to the geometric E(5)
critical point. We shall also explore the first-order critical
point in the phase transition from a spherical vibrator and an axially
symmetric deformed nucleus within the IBM.

\section*{Acknowledgments}

This work was supported in part by the Israel Science Foundation (A.L.)
and in part by the U.S. Department of Energy under contract
W-7405-ENG-36 (J.N.G).

\end{document}